\documentclass[amsmath,amssymb,twocolumn,showpacs,superscriptaddress,prl,longbibliography]{revtex4-1}

\usepackage{dcolumn}
\usepackage{bm}
\usepackage{graphicx}
\usepackage{epstopdf}
\usepackage{wrapfig}
\usepackage{hyperref}
\usepackage{array} 
\usepackage{listings}
\usepackage[para,online,flushleft]{threeparttablex}
\usepackage{booktabs,dcolumn}

\graphicspath{{.}{figures/}}

\usepackage{tikz}
\usetikzlibrary{arrows}
\usetikzlibrary{shadings}

\usepackage{textpos}
\usepackage{booktabs}
\usepackage{multirow,bigdelim}
\usepackage{float}
\usepackage{hyperref}

\usepackage{xcolor}
\definecolor{pastelgray}{rgb}{0.81, 0.81, 0.77}
\definecolor{beaublue}{rgb}{0.9, 0.9, 0.93}

\usepackage{siunitx}
\usepackage{todonotes}
\newcounter{mycomment}

\begin{document}

\preprint{RaF}

\title{\huge Spectroscopy of short-lived radioactive molecules: A sensitive laboratory for new physics}

\author{R.F. Garcia Ruiz}
\email{rgarciar@mit.edu}
\affiliation{CERN, CH-1211 Geneva 23, Switzerland}
\affiliation{Massachusetts Institute of Technology, Cambridge,
MA 02139, USA}
\author{R. Berger}
\email{robert.berger@uni-marburg.de }
\affiliation{Fachbereich Chemie, Philipps-Universit{\"a}t Marburg, Hans-Meerwein-Stra{\ss}e 4, 35032 Marburg, Germany}
\author{J.~Billowes}
\affiliation{School of Physics and Astronomy, The University of Manchester, Manchester M13 9PL, United Kingdom}
\author{C.L.~Binnersley}
\affiliation{School of Physics and Astronomy, The University of Manchester, Manchester M13 9PL, United Kingdom}
\author{M.L.~Bissell}
\affiliation{School of Physics and Astronomy, The University of Manchester, Manchester M13 9PL, United Kingdom}
\author{A.A. Breier} 
\affiliation{Laboratory for Astrophysics, Institute of Physics, University of Kassel, 34132 Kassel, Germany}
\author{A.J.~Brinson} 
\affiliation{Massachusetts Institute of Technology, Cambridge,
MA 02139, USA}
\author{K.~Chrysalidis}
\affiliation{CERN, CH-1211 Geneva 23, Switzerland}

\author{T.E.~Cocolios}
\affiliation{KU Leuven, Instituut voor Kern- en Stralingsfysica, B-3001 Leuven, Belgium}
\author{B.S.~Cooper}
\affiliation{School of Physics and Astronomy, The University of Manchester, Manchester M13 9PL, United Kingdom}
\author{K.T.~Flanagan}
\affiliation{School of Physics and Astronomy, The University of Manchester, Manchester M13 9PL, United Kingdom}
\affiliation{Photon Science Institute, The University of Manchester, Manchester M13 9PY, United Kingdom}
\author{T.F. Giesen}
\affiliation{Laboratory for Astrophysics, Institute of Physics, University of Kassel, 34132 Kassel, Germany}
\author{R.P.~de~Groote}
\affiliation{Department of Physics, University of Jyv\"askyl\"a, Survontie 9, Jyv\"askyl\"a, FI-40014, Finland}
\author{S. Franchoo}
\affiliation{Institut de Physique Nucleaire d'Orsay, F-91406 Orsay, France}
\author{F.P.~Gustafsson}
\affiliation{KU Leuven, Instituut voor Kern- en Stralingsfysica, B-3001 Leuven, Belgium}
\author{T.A. Isaev}
\affiliation{NRC "Kurchatov Institute"-PNPI, Gatchina, Leningrad district 188300, Russia}
\author{\'A.~Koszor\'us}
\affiliation{KU Leuven, Instituut voor Kern- en Stralingsfysica, B-3001 Leuven, Belgium}
\author{G.~Neyens}
\affiliation{CERN, CH-1211 Geneva 23, Switzerland}
\affiliation{KU Leuven, Instituut voor Kern- en Stralingsfysica, B-3001 Leuven, Belgium}

\author{H.A. Perrett}
\affiliation{School of Physics and Astronomy, The University of Manchester, Manchester M13 9PL, United Kingdom}
\author{C.M.~Ricketts}
\affiliation{School of Physics and Astronomy, The University of Manchester, Manchester M13 9PL, United Kingdom}
\author{S. Rothe}
\affiliation{CERN, CH-1211 Geneva 23, Switzerland}

\author{L.~Schweikhard}
\affiliation{Institut f\"ur Physik, Universit\"at Greifswald, Greifswald, Germany}
\author{A.R.~Vernon}
\affiliation{School of Physics and Astronomy, The University of Manchester, Manchester M13 9PL, United Kingdom}
\author{K.D.A.~Wendt}
\affiliation{Institut f\"ur Physik, Johannes Gutenberg-Universit\"at Mainz, D-55128 Mainz, Germany}
\author{F. Wienholtz}
\affiliation{CERN, CH-1211 Geneva 23, Switzerland}

\affiliation{Institut f\"ur Physik, Universit\"at Greifswald, D-17489 Greifswald, Germany}
\author{S.G.~Wilkins}
\affiliation{CERN, CH-1211 Geneva 23, Switzerland}

\author{X.F.~Yang}
\affiliation{School of Physics and State Key Laboratory of Nuclear Physics and Technology, Peking University, Beijing 100971, China}

\date{\today}

\maketitle

\textbf{The study of molecular systems provides exceptional opportunities for the exploration of the fundamental laws of nature and for the search for physics beyond the Standard Model of particle physics \cite{acme18,altuna18,berger:2019}. Measurements of molecules composed of naturally occurring nuclei have provided the most stringent upper bounds to the electron electric dipole moment (eEDM) to date \cite{acme18}, and offer a route to investigate the violation of fundamental symmetries with unprecedented sensitivity \cite{altuna18,Flamb14}. Radioactive molecules - where one or more of their atoms possesses a radioactive nucleus - can contain heavy and deformed nuclei, offering superior sensitivity for EDM measurements as well as for other parity- and time-reversal-violation effects \cite{Flamb19,isaev10}. Radium monofluoride, RaF, is of particular interest as it is predicted to have an appropriate electronic structure for direct laser cooling \cite{isaev10}, thus paving the way for high-precision studies on cold trapped molecules. Furthermore, some Ra isotopes are known to be pear shaped (octupole deformed) \cite{liam13}, thereby resulting in a large enhancement of their symmetry-violating nuclear moments \cite{auerbach96,kudashov14,Flamb19}. Until now, however, no experimental measurements of RaF have been performed, and their study is impeded by major experimental challenges, as no stable isotopes of radium exist. Here, we present a novel experimental approach to study short-lived radioactive molecules using the highly sensitive collinear resonance ionisation method. With this technique we have measured, for the first time, the energetically low-lying electronic states for each of the isotopically pure RaF molecules 
$^{223}$RaF, $^{224}$RaF, $^{225}$RaF, $^{226}$RaF, and $^{228}$RaF at the ISOLDE radioactive beam facility at CERN. 
Our results provide strong evidence of the existence of a suitable laser-cooling scheme for these molecules and constitute a pivotal step towards high-precision studies in these systems. Our findings open up new opportunities in the synthesis, manipulation and study of short-lived radioactive molecules, which will have a direct impact in many-body physics, astrophysics, nuclear structure, and fundamental physics research. 
}
\begin{figure*}[]
\includegraphics[scale=0.55,trim={2cm 0.5cm 0 0},clip]{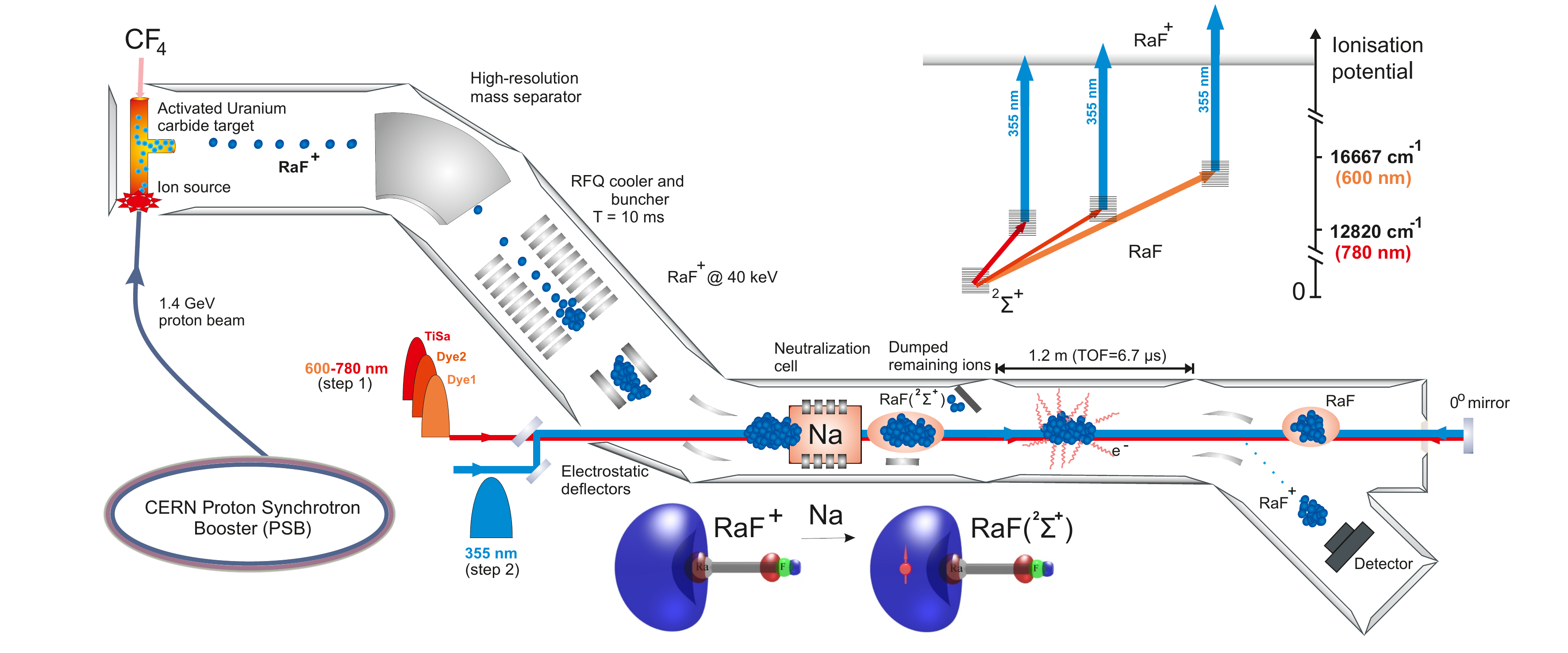}
\caption{\label{setup} 
Radioactive radium isotopes were created by impinging 1.4-GeV protons from the CERN Proton Synchrotron Booster (PSB) on a uranium carbide (UC$_x$) target. Radium monofluoride cations 
 (RaF$^+$) were produced by passing tetrafluoromethane (CF$_{4}$) gas through the activated UC$_x$ target at 1300~$^o$C. Molecular ions were extracted from the source, mass-selected, and injected into a helium-filled radio-frequency trap, where they were accumulated for 10 ms. Bunches of molecular ions were extracted and neutralised in-flight by charge exchange with neutral Na atoms. Neutral RaF molecules were overlapped with different laser beams in a collinear geometry. Resonantly reionised molecules were deflected onto a particle detector. The resonance ionisation scheme is shown on the top right. {Molecular orbitals are shown schematically.  Nuclear positions within the molecules are indicated by a grey sphere (Ra) and green sphere (F), whereas the sigma bond between the atoms
is indicated by a grey cylinder (see text for more details)}.}
\end{figure*}

Molecular systems are versatile laboratories to explore the atomic nucleus, as well as the properties and interactions
of the building blocks of matter (electrons and quarks) \cite{acme18,altuna18,safro18}. In molecules, electron-nucleon and nucleon-nucleon interactions are dominated by the electromagnetic and strong forces, respectively. The weak force, although much smaller in magnitude, can add measurable contributions to these interactions. Most importantly, the weak force is known to violate the symmetry with respect to spatial inversion of all particle coordinates (known as parity violation), giving rise to various intriguing phenomena. Some of these parity-violating effects have been measured with high accuracy for only a few atomic systems \cite{Wood97,Anty19,safro18}, contributing to the most stringent low-energy tests of the Standard Model. In certain molecules, effects resulting from both parity violation (P-odd) and time-reversal violation (T-odd) are {significantly enhanced with respect to atomic systems \cite{auerbach96,kudashov14,safro18,Flamb19}}, offering the means to explore unknown aspects of the weak and strong forces.  As the strengths of these interactions scale with the atomic number, the nuclear spin and nuclear deformation, molecular compounds of heavy radioactive nuclei are predicted to exhibit {unprecedented sensitivity, with an enhancement of more than two orders of magnitude} for both P-odd and P,T-odd effects \cite{auerbach96,isaev10,kudashov14,gaul18,Flamb19}.\\ However, the experimental knowledge of radioactive molecules is scarce \cite{For17}, and quantum chemistry calculations often constitute the only source of information.  Molecules possess complex quantum level structures, which renders spectroscopy of their structure considerably more challenging when compared to atoms. Moreover, major additional experimental challenges must be overcome to study molecules containing heavy and deformed short-lived nuclei. These radioactive nuclei do not occur naturally and must therefore be produced artificially at specialised facilities such as the Isotope Separator On-line Device (ISOLDE) at CERN. As comparatively little is known about the properties and chemical bonds of radioactive molecules, the techniques to synthesise and manipulate them are not fully understood \cite{For17}. Furthermore, molecules containing short-lived isotopes can only be produced in quantities smaller than $10^{-8}$ grams (typically with rates of less than 10$^6$ particles/s). Thus, spectroscopic
studies require particularly sensitive experimental
techniques adapted to the properties of radioactive ion
beams and conditions present at radioactive beam facilities. Here, we present a new approach for performing laser spectroscopy of short-lived radioactive molecules, providing the first spectroscopic information of radium monofluoride (RaF). To our knowledge, this is the first laser spectroscopy study ever performed on
with a molecule containing short-lived isotopes. 

Since the theoretical suggestion for direct cooling of monofluoride molecules with lasers \cite{DiRosa:04} has been experimentally demonstrated \cite{shuman10}, a great deal of experimental and theoretical attention has been focussed on molecular systems with similar structures \cite{hudson11,barry14,truppe17,lim18,anderegg18}. The laser cooling of polyatomic molecules has been proposed and criteria to identify suitable candidates have been outlined \cite{isaev16}. This has triggered a wealth of studies on the opportunities resulting from laser cooling techniques in molecular physics \cite{isaev18}. In contrast to other heavy-atom molecules, RaF is predicted to have highly closed excitation and re-emission optical cycles, which would make it ideal for laser cooling and trapping \cite{isaev10}. Moreover, due to the recently discovered pear-shaped nuclear deformation of certain Ra isotopes \cite{liam13}, the interactions of the electrons with the P-odd nuclear anapole moment as well as with the P,T-odd nuclear Schiff and magnetic quadrupole moments are predicted to be enhanced {by more than two orders of magnitude} \cite{isaev13,Flamb14,Flamb19,gaul18}. Hence, these molecules could provide a unique laboratory to measure these symmetry-violating nuclear moments.

Figure~\ref{setup} shows a diagram of the experimental setup used to produce and study the RaF molecules. 
 As a first step, radium isotopes were produced by evaporation of an irradiated target (see section Methods - Production of RaF molecules). RaF$^{+}$ molecular ions were then formed upon injection of CF$_{4}$ gas into the target material. Molecular ions were extracted from the ion source by applying an electrostatic field, and molecules containing one specific radium isotope were selected with a high-resolution magnetic mass separator ($\Delta m/m \sim 1/5000$). 
The ions were collisionally cooled in a radio-frequency quadrupole (RFQ) trap filled with helium gas at room temperature. After 10~ms of cooling time, bunches of RaF$^{+}$ with a 4-\si{\micro\second} temporal width were released and accelerated to 39998(1)~eV, before entering into the Collinear Resonance Ionisation Spectroscopy (CRIS) setup \cite{Flan13,garcia18}. \\At the CRIS beam line, the ions were first neutralised in-flight by passing through a collision cell filled with sodium vapour, inducing charge-exchange according to the reaction RaF$^+$ $+$  Na $\rightarrow$ RaF $+$ Na$^{+}$. As the ionisation energy of RaF is estimated to be close to that of Na (5.14~eV) \cite{isaev13b}, the neutralisation reaction dominantly populates the RaF 
 $X\, ^2\Sigma^+$ electronic ground state. {Molecular orbitals are shown schematically in Figure \ref{setup}. The lowest unoccupied molecular orbital in RaF$^+$, which is
   essentially of non-bonding character, becomes occupied by an unpaired
   electron (symbolised in Figure 1 by a red sphere together with an arrow
   representing the electron spin) upon neutralisation. This is shown schematically as an isodensity surface, with lobes in slightly transparent blue and transparent red indicating different relative phases of the single-electron wavefunction.}

 After the charge-exchange reaction, non-neutralised RaF$^+$ ions were deflected out of the beam, and the remaining bunch of neutral RaF molecules was overlapped in time and space by several (pulsed) laser beams in a collinear arrangement, along the ultra-high-vacuum ($10^{-10}$~mbar) interaction region of 1.2-m length.
Laser pulses (step~1) of tunable wavelength were used to resonantly excite the transition of interest, and a high-power 355-nm laser pulse (step~2) was used to subsequently ionise the excited RaF molecules into RaF$^+$ (see inset of Figure \ref{setup}). The resonantly ionised molecules were then separated from the non-ionised molecules by deflecting the ions onto a particle detector. When the excitation laser is on resonance with a transition in the molecule (step 1 in Figure \ref{setup}), the second laser pulse ionises the molecule, producing a signal at the detector.  Molecular excitation spectra were obtained by monitoring the ion counts as a function of the wavenumber of the first laser.

As only theoretical predictions were available for the excitation energies of RaF, finding the transition experimentally required scanning a large wavelength range ($>$1000~cm$^{-1}$). The prediction for the  
$A \,^2\Pi_{1/2} - X \, ^2\Sigma^+$ (0,0) transition, for example, was 13300~cm$^{-1}$ with an accuracy estimated to be within 1200~cm$^{-1}$ \cite{isaev10,isaev13}. Given the bandwidths of laser available ($<$ 0.3 cm$^{-1}$), the scan of such a large wavelength region on samples produced at rates below 10$^6$ molecules/s represents a significant experimental challenge. In order to optimize the search of molecular transitions, three broadband lasers were scanned simultaneously (see section Methods - Laser setup), and a zero degree mirror at the end of the beam line was used to reflect the laser light anti-collinearly with respect to the travelling direction of the RaF bunch. Thus, each of the three scanning lasers covered two different wavenumber regions separated by 15.7 cm$^{-1}$ in the molecular rest frame due to the Doppler shift present for the fast RaF molecules (see section Methods). The possibility of using multiple high-power broadband lasers to scan large wavenumber ranges both 
 collinearly 
   and anti-collinearly is a marked advantage of the CRIS technique \cite{garcia18} compared with optical detection techniques, which would suffer from a large background due to the presence of intense photon densities delivered close to the photodetectors.

\begin{figure*}[]
\includegraphics[scale=0.27,trim={1.0cm 0cm 0 0},clip]{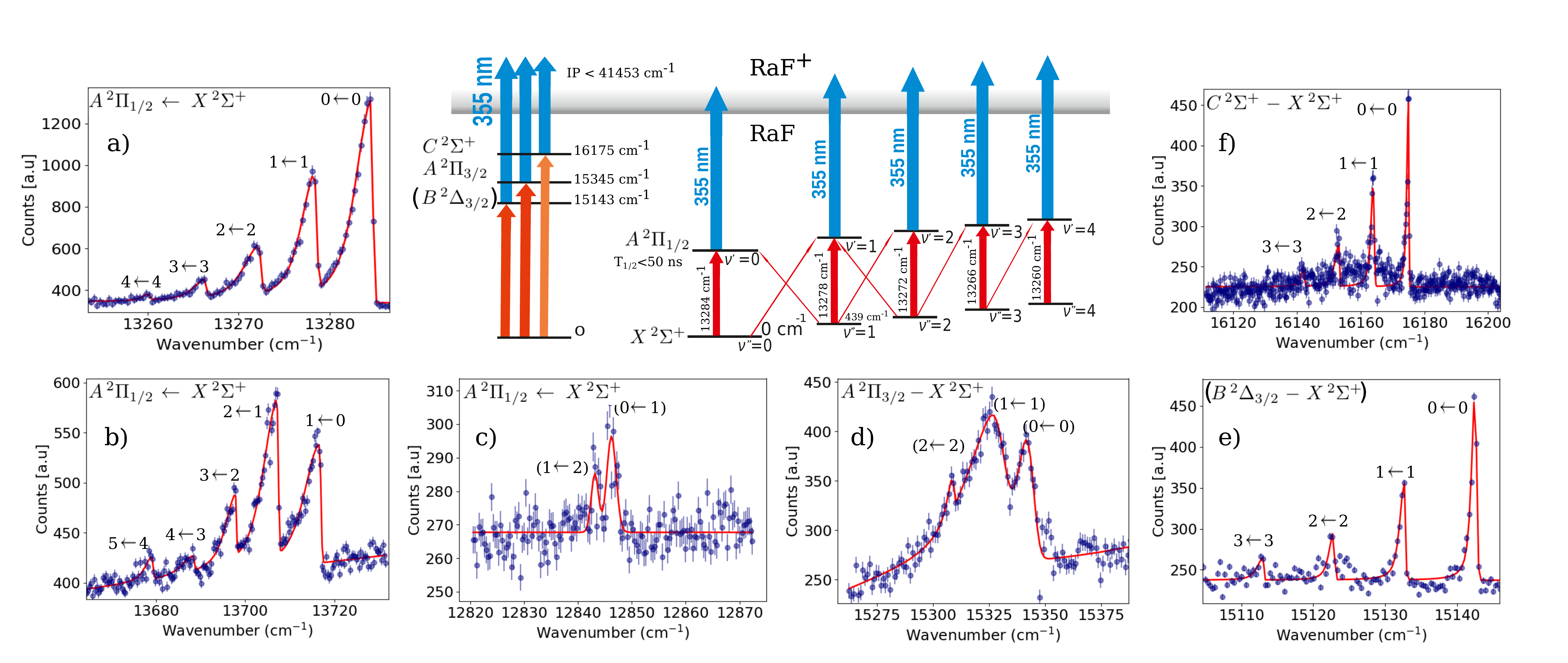}
\vspace{-0.8cm}
\caption{\label{RaFscheme} 
Measured vibronic spectra for $^{226}$RaF. Wavenumbers in the spectra are in the rest frame of the molecule. The counts on the particle detector were measured as a function of the laser wavenumber of the resonant step. A fixed wavelength (355 nm) was used for the ionisation step. Panel a) shows the observed peaks corresponding to the vibronic spectra of the $\Delta v$=0 band system of $v''$=0, 1, ..4, scanned by the grating Ti:Sapphire laser. The pulsed dye laser was used to scan electronic transitions in different wavelength ranges: b) the $\Delta v$=+1 band system of the $A \,^2\Pi_{1/2}$ $\leftarrow$ $X \,^2\Sigma^+$ transition with $v''$=0, 1, ..4, and c) the ($v', v''$) = (0,1) and (1,2) band. The corresponding transitions to other electronic states are shown in panels: d) $A \,^2\Pi_{3/2}$ $\leftarrow$ $X \,^2\Sigma^+$, e) $B \,^2\Delta_{3/2}$ $\leftarrow$ $X \,^2\Sigma^+$ (tentatively assigned), and f) $C \,^2\Sigma^+$ $\leftarrow$ $ X \,^2\Sigma^+$. The shape of the spectra is due to population distribution of different rotational states. The solid lines show the fit with skewed Voigt profiles. Three essential properties for laser cooling of RaF molecules were identified: i) the short lifetime of the excited states $^2\Pi_{1/2}$ ($T_{1/2}<$ 50 ns) will allow for the application of strong optical forces, ii) dominant diagonal transitions, ($\Delta v$=0)/($\Delta v$=$\pm 1)>0.97$, indicate a large diagonal Franck-Condon factor, and iii) the expected low-lying electronic states $B \,^{2}\Delta_{3/2}$, $A \,^2\Pi_{3/2}$, and $C \,\Sigma^+$ were found to be above the $A \,^2\Pi_{1/2}$ states, which will allow efficient optical cooling cycles to be applied.}
\end{figure*}

The predicted region for the $A \,^2\Pi_{1/2} \leftarrow X\,^2\Sigma^+$  transition was scanned (with the three lasers for the first step simultaneously and both collinearly and anti-collinearly) at a speed of 0.06~cm$^{-1}$/s, covering a 1000~cm$^{-1}$ range in only 5 hours.
 After a few hours of scanning on a beam of $^{226}$RaF, a clear sequence of vibronic absorption signals was recorded. The measured spectrum  assigned to the  ($v',v''$) vibrational transitions (0,0), (1,1), (2,2), (3,3) and (4,4) of the $A \,^2\Pi_{1/2} - X\,^2\Sigma^+$ band system is shown in Figure~\ref{RaFscheme}a. 
Weaker band structures found around 440 cm$^{-1}$ s higher and lower with respect to the (0,0) band were assigned to the  $\Delta v=\pm1$ transitions ($v',v''$) = (1,0), (1,2), (2,1), (3,2), (4,3), (5,4) and (0,1), (1,2), respectively (Figures~\ref{RaFscheme}b and \ref{RaFscheme}c). {The quantum number assignment for $\Delta v=-1$ is tentative, due to the highly dense structure of overlapping vibronic bands.}

{In addition to the $A \,^2\Pi_{1/2} - X \, ^2\Sigma^+$  band system, we found spectroscopic signatures of electronic transitions to higher-lying states. Some examples
of recorded spectra are shown in Figures \ref{RaFscheme}d-f, along with the energy-level scheme. We assign the observed transitions as follows: The band system around $15325~\mathrm{cm}^{-1}$ (Figure \ref{RaFscheme}d) is attributed
to the $A \,{}^{2}\Pi_{3/2} - X \,{}^{2}\Sigma^+$ transition due to the complex
rovibrational structure expected to arise from intense satellites
that are possible in these transitions. As the
bands are comparatively strong, they are assigned to the $\Delta v=0$ band
system. Whereas the individual assignments to vibrational transitions must
be considered tentative as per the congested structure of the
Franck--Condon profile, the $\Delta v=0$ assignment is supported by the fact that no additional structure was located within a $\pm$ 400 cm-1 region. The band system located around 15142.7 cm$^{-1}$ (Figure \ref{RaFscheme}e)
is tentatively assigned to the $B \,{}^{2}\Delta_{3/2} - X \, {}^{2}\Sigma^+$
transition by virtue of the good agreement with the computed excitation
energies to the $\Omega=3/2$ state of mixed $\Delta/\Pi$ character \cite{isaev10,isaev13}. This
mixing provides intensity to the one-photon transition from a $\Sigma$
state into the $\Delta$ manifold. The computed Born-Oppenheimer potentials
for this $\Omega=3/2$ state and the electronic ground state are, however,
highly parallel, which would suggest a sparser Franck-Condon profile than was observed experimentally. However, we note that the related $B \,{}^{2}\Delta_{3/2} - X \,{}^{2}\Sigma^+$transition in BaH and BaD were reported to have a perturbed character due to mixing between electronic levels \cite{bernard:1989}. Thus, in the
present case, a vibrational profile that is richer than expected from
adiabatic potentials can not be ruled out a priori. The band system with
origin at $16175.2~\mathrm{cm}^{-1}$ (Figure \ref{RaFscheme}f) is assigned to the
$C \,{}^{2}\Sigma^+ - X \,{}^{2}\Sigma^+ $ transition based on the observed
Franck--Condon profile, which is in good agreement with the computed
harmonic vibrational energy spacings as well as the expected intensity
distribution and is in a wavenumber region that is only slightly lower than
predicted \cite{isaev10,isaev13}.} All measured and assigned vibronic bands of the four electronic transitions 
are listed in Table \ref{tab:table1}. 

The measured $A \,^2\Pi_{1/2} - X \, ^2\Sigma^+$ (0,0) band center, $\tilde{\mathcal{T}}_e~=~13287.8(1)$ cm$^{-1}$ is in excellent agreement with the {\it ab initio} calculated value of 13300(1200) cm$^{-1}$ \cite{isaev13}.  
In accordance with theoretical predictions \cite{isaev10}, we found vibronic transitions with $\Delta v$=0 to be much stronger than those of $\Delta v$=$\pm 1$.  
 For most of the measurements the power density used for the resonant step was 100(5)~\si{\micro\joule}\,cm$^{-2}$ per pulse as measured at the entry window of the CRIS beam line. Reducing the power by 50~$\%$ did not significantly reduced the resonant ionisation rate, indicating that these transitions were measured well above saturation.
 The much weaker vibrational transitions with $\Delta v$=$\pm1$
 were scanned with a pulsed dye laser of 500(5)~\si{\micro\joule}\,cm$^{-2}$ power density per pulse (bandwidth of 0.1 cm$^{-1}$).
 As the $\Delta v=\pm1$ transitions 
 were measured well above saturation and with laser beams of different characteristics, a precise
 estimation 
 of the Franck-Condon (FC) factors 
 could not 
 be obtained. Instead, an upper limit of 0.03  
for the peak intensity ratio of $I(0,1) / I(0,0)$  was derived, 
indicating 
highly diagonal FC factors, an essential property for laser cooling \cite{isaev10}.  \\
By measuring the resonant ionisation rate for different time delays between the excitation and ionisation laser pulses, an upper limit for the lifetime of the excited state $^2\Pi_{1/2}$ ($v'$=0), $T_{1/2} \leq$ 50~ns, was obtained. The measurements were performed with the wavenumber of the resonant laser fixed at the resonance value of the transition ($v',v''$) = (0,0).
The resonant ionisation rate dropped by more than 70 \% for time delays above 50~ns. This short lifetime corresponds to a large spontaneous decay rate ($>$ 2$\times$10$^{7}$~s$^{-1}$), which would allow for the application of strong optical forces for laser cooling. An additional concern for the suitability of laser cooling is related to the existence of metastable states lying energetically below the $^2\Pi_{1/2}$ level, which could prevent the application of a closed optical cooling loop, a major problem encountered for BaF \cite{isaev10}. In contrast to BaF, all other predicted electronic states ($^2\Pi_{3/2}$, $^2\Delta_{3/2}$ and $^2\Sigma$) in RaF were found to be energetically above the  $^2\Pi_{1/2}$ state. This demonstrates a marked advantage of RaF as its electronic structure will allow for efficient optical-cooling cycles.

\begin{table}
\caption{Measured vibronic transitions of $^{226}$RaF from the $X\,^2\Sigma^+$ electronic ground state to excited $A \,^2\Pi$ and $B \,^2\Delta$ states. The values indicate the band head positions. Combined statistical and systematic uncertainties are included in parentheses.}
\label{tab:table1}
\begin{ruledtabular}
\begin{tabular}{ c c d }
  Transition & $v^{\prime}$  $\leftarrow$  $v^{\prime\prime}$ & \Delta\tilde{\nu}/\mathrm{cm}^{-1}\\ \colrule\\
A $\,^2\Pi_{1/2}$ $\leftarrow$ X $\,^2\Sigma^+ $& $0 - 0$ & 13284.7(5)    \\
  &  $1 - 1$  &13278.5(5)     \\
  &  $2 - 2$  &13272.4(5)     \\
  &  $3 - 3$  &13266.4(10)      \\
  &  $4 - 4$  &13260.2(10)      \\ 
  
  &  $1 - 0 $  &13716.9(5)      \\ 
  &  $2 - 1 $  &13707.4(5)      \\
  &  $3 - 2 $  &13698.0(5)      \\ 
  &  $4 - 3 $  &13688.6(10)      \\ 
  &  $5 - 4 $  &13679.4(10)      \\ 
  
  &  ($0 - 1 $)  &12846.3(10)      \\  
  &  ($1 - 2 $)  &12843.1(10)  \\ \colrule\\
 ($B \,^{2}\Delta_{3/2}$  $\leftarrow$ $X \,^2\Sigma^+$) & $0 - 0$& 15142.7(5)  \\
  &  $1 - 1$  &15132.8(10)      \\
  &  $2 - 2$  &15123.0(10)      \\
  &  $3 - 3$  &15113.2(10)      \\ \colrule\\
 $A \,^{2}\Pi_{3/2}$  $\leftarrow$ $X \,^2\Sigma^+$ & ($0 - 0$)& 15344.6(50)  \\
  &  ($1 - 1$)  &15325.0(80) \\ 
  &  ($2 - 2$)  &15309.4(100) \\   \colrule\\
 
$C \,^{2}\Sigma^+$ $\leftarrow$ $X \,^2\Sigma^+$ & $0 - 0$& 16175.2(5)  \\
  &  $1 - 1$  &16164.2(5)      \\
  &  $2 - 2$  &16153.4(5)      \\
  &  $3 - 3$  &16142.4(10)      \\ 
  
\end{tabular}
\end{ruledtabular}
\end{table}

From combination differences of energetically low-lying vibronic transitions in the band system $A \,^2\Pi_{1/2} - {X \,^2\Sigma^+}$  we have derived experimental values for the  harmonic frequency, $\tilde{\omega}_e$, and the dissociation energies, $\tilde{\mathcal{D}}_e$, using a Morse potential approximation.  
\begin{table}[htbp]
	\caption{ $^{226}$RaF Morse potential parameters for $X^2\Sigma^{+}$ electronic ground and $A \,^2\Pi_{1/2}$ excited states.} 
\begin{ruledtabular}
	\begin{tabular}{lrr}
		 Parameter &	$X^2\Sigma^+$/cm$^{-1}$ &  $A^2\Pi_{1/2}$/cm$^{-1}$ \\		\colrule
		$\tilde{\omega}_e$							& 441.8(1)		& 435.5(1) \\
		$\tilde{\mathcal{D}}_e$/$10^{4}$ 	&2.92(5)& 2.90(3)\\
	\end{tabular}%
\end{ruledtabular}
 	\label{tab:MorsePara}
\end{table}
Results are given in Table ~\ref{tab:MorsePara}, and further details of the analysis can be found in the section Methods - Spectroscopic analysis.\\

\begin{figure}[]
\includegraphics[scale=0.28]{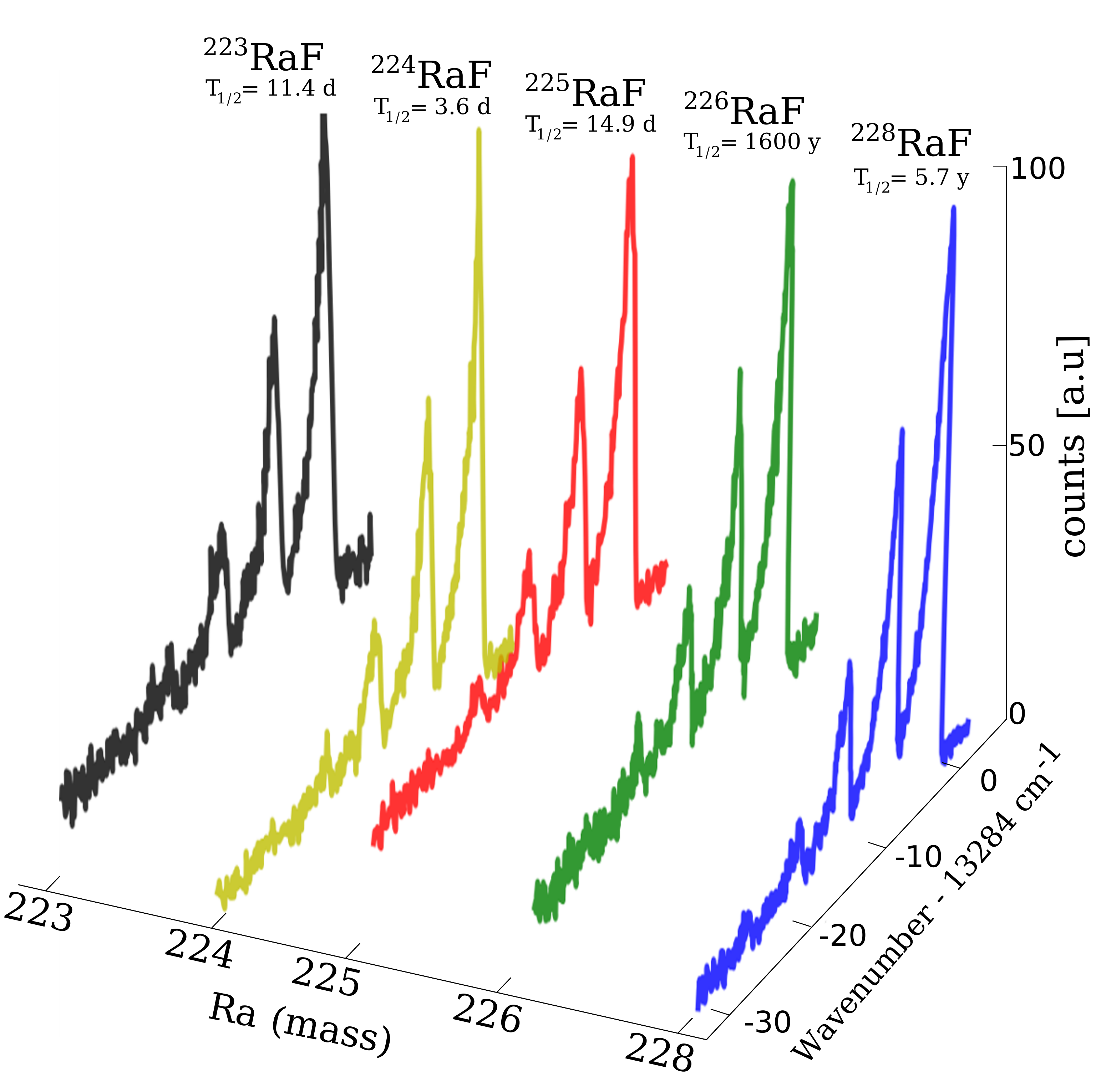}
\caption{\label{spectra} Measurements of the vibronic absorption spectra  $A \,^2\Pi_{1/2}$  $\leftarrow$ $X \,^2\Sigma^{+}$ of the isotopologues $^{223}$RaF, $^{224}$RaF, $^{225}$RaF, $^{226}$RaF, and $^{228}$RaF. }
\end{figure}

Furthermore, the $A \,^2\Pi_{1/2}\leftarrow \,X \,^2\Sigma^+$ vibronic spectra of the short-lived isotopologues $^{223}$RaF, $^{224}$RaF, $^{225}$RaF, $^{226}$RaF and $^{228}$RaF, were measured (Figure \ref{spectra}). All vibrational transitions were clearly observed, including those of the molecule with the shortest-lived Ra isotope studied, $^{224}$RaF ($T_{1/2}= 3.6$ d). An on-line irradiation of the target material will allow molecules containing isotopes with lifetimes as short as a few milliseconds to be studied. 
The main limitation is dictated by the release from the target and the trapping time spent in the radio-frequency trap ($>$ 5 ms). Future high-resolution measurements
will enable studies of nuclear structure changes resulting
from different isotopes and nuclear spins.

{In summary, this letter presents a new experimental approach to perform laser spectroscopy studies of molecules containing radioactive nuclei. The high sensitivity of the technique allowed the study of RaF molecules produced at rates lower than $10^{6}$ molecules/s. Our results have established the energetically low-lying electronic structure of these molecules, and constitute a pivotal step towards precision measurements in this system. Three essential properties for the laser cooling of RaF molecules were identified: i) the short lifetime of the excited states $A \,^2\Pi_{1/2}$ ($T_{1/2}<$ 50 ns) will allow for the application of strong optical forces, ii) dominant diagonal transitions, indicating a large diagonal Franck-Condon factor, and iii) the expected low-lying electronic states $B \,^{2}\Delta_{3/2}$, $A \,^2\Pi_{3/2}$, and $C \,^2\Sigma^+$ states were found to be above the $A \,^2\Pi_{1/2}$ states, which will allow efficient optical cooling cycles to be applied.}

The ability to produce, mass-select, and spectroscopically study radioactive molecules is expected to profoundly impact several fields of research. For example, astronomical observations have recently suggested the presence of the long-lived radioactive molecule $^{26}$AlF, based on theoretical predictions of its molecular structure \cite{kami18}. Our experimental method will enable spectroscopy studies of this radioactive molecule. These studies can be extended to other radioactive compounds of astrophysical interest (e.g. $^{14}$CO).\\
 The new experimental technique can also be employed for the laser spectroscopy of a wide variety of neutral molecules and molecular ions, including those with very short-lived isotopes ($\leq$ 1 day). Radioactive molecules can be specifically tailored to enhance the sensitivity to parity- and time-reversal-violating effects by introducing heavy and deformed nuclei. Moreover, by systematically replacing their constituent nuclei with different isotopes of the same element, both nuclear-spin-independent and nuclear-spin-dependent effects can be comprehensively studied. Thus, electroweak interactions and the influences from possible P,T-odd nuclear moments can be investigated as a function of the nuclear spin quantum number, $I$, of their constituent isotopes. In the case of RaF molecules, for example, Ra isotopes with a variety of nuclear spins are available, such as the $I=0$ nuclei $^{224}$Ra, $^{226}$Ra, $^{228}$Ra, the $I=1/2$ nuclei $^{213}$Ra, $^{225}$Ra, the $I=3/2$ nuclei $^{223}$Ra,  $^{227}$Ra, and the $I=5/2$ nuclei $^{229}$Ra. Future high-precision measurements will allow for the study of nuclear structure effects, as well as still unexplored subatomic properties such as the P-odd T-even nuclear anapole moments, the P,T-odd nuclear Schiff moment and magnetic quadrupole moments.\\  Currently, most of our knowledge of nuclear ground-state electromagnetic properties of unstable nuclei has been obtained from the study of radioactive atoms \cite{garciaruiz16,campbell16,Marsh18}, but very little is known about the nuclear weak structure.  Further precision studies of radioactive molecules will offer a new window in the exploration of the atomic nucleus and its fundamental constituents. Our results open up new opportunities to study fundamental physics, nuclear structure, and quantum chemistry of the heaviest elements. Thus, we expect that these findings will motivate new avenues of research at the increasingly capable radioactive-ion-beam facilities around the world.

\textit{Acknowledgements.}---
This work was supported by the ERC Consolidator Grant No.648381 (FNPMLS); 
Deutsche Forschungsgemeinschaft (DFG, German Research Foundation) -- Projektnummer 328961117 -- SFB 1319 ELCH;
STFC grants ST/L005794/1 and ST/L005786/1 and Ernest Rutherford Grant No. ST/L002868/1; projects from FWO-Vlaanderen GOA 15/010 from KU Leuven and BriX IAP Research Program No. P7/12; the European Unions Grant Agreement 654002 (ENSAR2); the Russian Science Foundation under grant N 18-12-00227; the BMBF grants 05P15HGCIA and 05P18HGCIA. We thank J. P. Ramos and T. Stora for their support in the production of RaF molecules. We would also like to thank the ISOLDE technical group for their support and assistance. We thank D. Budker for comments and suggestions as well as Alexander Petrov for discussions on $\Delta$ states. R.B. acknowledges I. Tietje for early discussions on various experiments at CERN and is indebted to A. Welker for sharing knowledge on isotope production and separation as well as for initial discussions of the RaF studies. R.B. acknowledges discussions with
   K. Gaul on molecular properties and with D. Andrae on finite nuclear size effects. R.B. and T.A.I acknowledge S. Hoekstra and L. Willmann for early discussions on production of RaF. T.A.I. is grateful to A. Zaitsevskii for discussions on the coupled-cluster method.

\textit{Author contributions}---
R.F.G.R. led the experimental part and R.B. led the theoretical support. R.F.G.R., R.B., C.B., M.L.B., K.C., B.S.C., K.T.F., R.P.G, S.F., F.P.G., A.K., H.A.P., C.M.R., S.R., A.R.V., F.W. and S.G.W performed  the  experiment.   R.F.G.R., R.B., A.A.B., A.J.B. and T.F.G. performed  the data analysis. R.F.G.R. prepared the figures. R.B. and T.A.I performed theoretical calculations that motivated the experimental proposal and analysis of the results. R.F.G.R, R.B. prepared the initial draft of the manuscript with input from A.A.B., A.J.B., K.T.F., T.F.G., T.A.I., G.N. and S.G.W.  All authors discussed the results and contributed to the manuscript at different stages.

\textbf{Data availability.}
 All the relevant data supporting the findings of these studies are available from the corresponding author upon request.

\bibliography{RaF}

\section{methods}
\textbf{Production of RaF molecules:}  Radium isotopes were produced 33 days before the laser-spectroscopy measurements by impinging 1.4-GeV protons on the cold uranium carbide target material. The target was exposed to pulses of $10^{13}$ protons/pulse over a period of $2$~days. After irradiation with a total of 8$\times 10^{17}$ protons, the target was kept in a sealed chamber filled with Ar gas. After day $33$, the target was connected to the High-Resolution Separactor (HRS) front end at ISOLDE. FLUKA \cite{boh14} simulations predicted $2 \times 10^{13}$ atoms of $^{226}$Ra in the target material (7.5 $\times$ $10^{-9}$ grams), following proton irradiation. The target was pumped down to pressures below $10^{-5}$~mbar, and the target holder and ion source were gradually heated up to about 1300~$^\text{o}$C. A leak valve attached to the target was used to inject CF$_{4}$ into the target environment. RaF molecules were formed by reactive collisions of CF$_{4}$ molecules with Ra atoms \cite{isaev13b} present inside the irradiated target material.

\begin{figure}[t]
\includegraphics[scale=0.4,trim={0.5cm 0.5cm 0 0},clip]{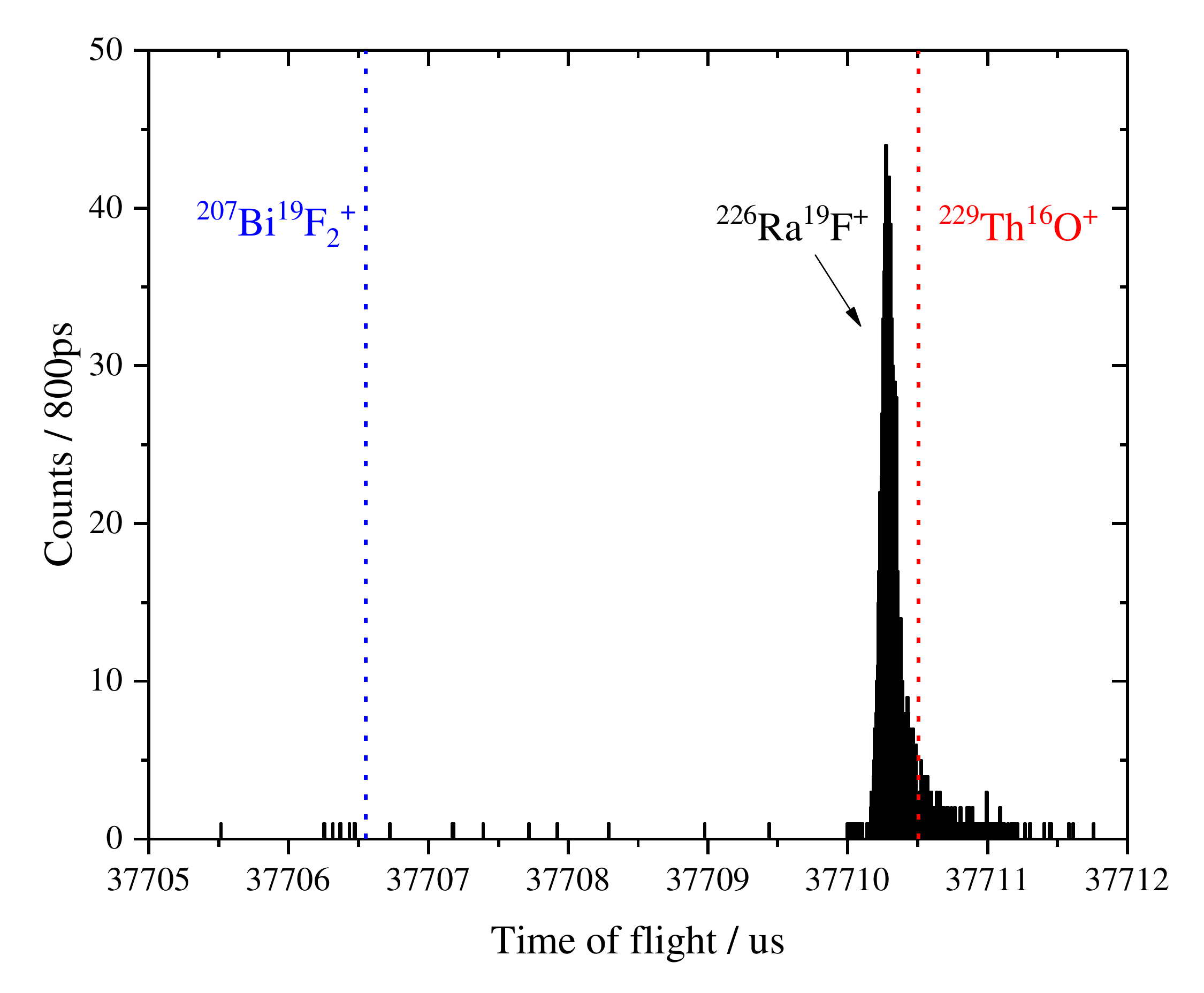}
\caption{\label{mrtof} Time-of-flight spectrum of the $^{226}$RaF$^{+}$($A$=245) beam as delivered from ISOLDE after 1000 revolutions in the MR-ToF MS. A mass resolving power $1.7 \times 10^5$ was achieved which allowed the isobaric beam composition to be analysed. Only $^{226}$Ra$^{19}$F$^{+}$ ions were detected. The position of the most likely accompanying ions are highlighted by dotted vertical lines.} 
\end{figure}

The $^{226}$RaF$^{+}$($A$=245) beam extracted from the ISOLDE target unit was sent to the ISOLTRAP setup \cite{kre13}, where the molecular ions were captured, cooled and bunched by a RFQ trap and subsequently analysed using a multi-reflection time-of-flight mass spectrometer (MR-ToF MS) \cite{wol13}. A measured mass spectrum is shown in Figure \ref{mrtof}. After 1000 revolutions in the device, a mass resolving power ($R=m/\Delta m$) of $1.7 \times 10^5$ was achieved, which allowed the isobaric beam composition to be analysed. The only mass peak detected was identified as the signal of $^{226}$Ra$^{19}$F$^{+}$, confirming the purity of the beam from ISOLDE. 

\textbf{Laser setup.} The resonant ionisation schemes used for the study of RaF molecules are shown in Fig. \ref{setup}.  Three different laser systems were prepared to cover the scanning range from 12800~cm$^{-1}$ to 13800~cm$^{-1}$: i) a dye laser system (Dye1: Spectron Spectrolase 4000) provided pulses of 100(5)~\si{\micro\joule} with a linewidth of 10~GHz  (0.3~cm$^{-1}$).  ii) a dye laser (Dye2: Sirah Cobra) with a narrower linewidth of 2.5~GHz (0.09~cm$^{-1}$) produced pulses of similar energy. The lasers were loaded with either Styryl 8  or DCM dyes to provide wavenumber ranges 12800-14000~cm$^{-1}$ and 15150-16600~cm$^{-1}$, respectively. Both dye lasers were pumped by 532-nm pulses at 100~Hz, obtained from two different heads of a twin-head Nd:YAG laser (Litron LPY 601 50-100 PIV). iii) a grating Ti:Sapphire laser system with a linewidth of 2~GHz (≡0.07~cm$^{-1}$) produced pulses of 20(1)~\si{\micro\joule}, pumped by 532-nm 
pulses 
 at 1~kHz from a Nd:YAG laser (LDP-100MQ LEE laser). The non-resonant ionisation step was obtained by 355-nm pulses of 30~mJ at 100~Hz, produced by the third-harmonic output of a high-power Nd:YAG laser (Litron TRLi).
\\The release of the ion bunch was synchronised with the laser 
pulses 
 by triggering the flash-lamps and Q-switch of the 
 pulsed 
  lasers with a  Quantum Composers 9528 digital delay pulse generator. \\
Dye laser wavelengths were measured with a WS-6 wavelength meter. Ti:Sapphire laser wavelengths were measured by a WSU-2 HighFinesse wavelength meter calibrated by measuring a reference wavelength provided by a stabilised diode laser (TopticaDLC DL PRO 780).

\textbf{Collinear and anti-collinear excitation.} For a molecule travelling at velocity $v$, the laser wavenumber in the laboratory frame, $\tilde{\nu}_0$, is related to the wavenumber in the molecule rest frame, $\tilde{\nu}$, by the expression $\tilde{\nu}=\dfrac{1+ \beta \cos \theta}{\sqrt{1-\beta^2}} \tilde{\nu}_{0}$, with $\beta=v/c$ and $\theta$ being the angle between the direction of the laser beam and the velocity of the molecule. For RaF molecules at 39998(1)~eV ($v~\sim~ 0.18$~m/$\mu$s), a difference of 15.7~cm$^{-1}$ is obtained between the laser pulse sent out collinearly ($\cos \theta=1$) and anti-collinearly ($\cos \theta=-1$) with respect to the direction of the molecule's velocity.

\textbf{Spectroscopic analysis.}
The vibrational transitions in Fig. 2 show asymmetric line profiles with a maximum located towards higher wavenumbers. Since the band centers cannot be determined directly from the measured line profiles, we used the wavenumber positions of the maxima in our data analysis. Table \ref{tab:table1} lists the maximum peak positions and estimated uncertainties are given in parentheses.  The wavenumber difference, $\Delta \tilde{\nu}$, of vibrational levels in the electronic $^2\Sigma^+$ ground state and in the $^2\Pi_{1/2}$ excited state were derived from combination differences of the recorded $^{226}$RaF spectra (see Table \ref{tab:combD}).

\begin{table}[htbp]
	\caption{$^{226}$RaF vibrational transitions in the electronic $X^2\Sigma^+$ ground state and $A^2\Pi_{1/2}$ excited state derived from combination differences.  Uncertainties are given in parentheses.}
\begin{ruledtabular}
	 	\begin{tabular}{cll}
\toprule
$v^\prime$ $\leftarrow$ $v^{\prime\prime}$ & $^2\Sigma^+$ $\Delta$ $\tilde{\nu}/\text{cm}^{-1}$ & $^2\Pi_{1/2}$ $\Delta\tilde{\nu}/\text{cm}^{-1}$
\\ \hline
		\midrule
		1\,-\,0 & 438.4(7)& 432.2(7)\\
		2\,-\,1 & 435.0(7)& 428.9(7)\\
		2\,-\,1 & 435.4(11)& 429.3(11)\\
		3\,-\,2 & 431.6(11)& 425.6(7)\\
		4\,-\,3 & 428.4(14)& 422.2(14)\\
		5\,-\,4 &$\cdots$& 419.2(14)\\
		\bottomrule
	\end{tabular}%
	\end{ruledtabular}
	\label{tab:combD}
\end{table}
In our analysis we used vibrational energy terms $E_v/(hc)$ of a Morse potential according to: 
\begin{equation}
E_v/(hc)= \tilde{\omega}_e \left(v+\frac{1}{2}\right)-\frac{\tilde{\omega}^2_e}{4 \tilde{\mathcal{D}}_e}\left(v+\frac{1}{2}\right)^2.
\end{equation}
Energy level differences 
\begin{equation}
	(E_{v+1}-E_v)/{(hc)}= \tilde{\omega}_e - \frac{\tilde{\omega}^2_e}{2\tilde{\mathcal{D}}_e}\left(v+1\right),
	\label{eq:spacing}
\end{equation}
were used to derive the Morse potential parameters $\tilde{\omega}_e$ and $\tilde{\mathcal{D}}_e$ from a least-squares fit analysis.  The derived energy level differences are given in Table \,\ref{tab:combD}, whereas Table \,\ref{tab:MorsePara} contains the molecular parameters from the fit. The harmonic vibration frequencies $\tilde{\omega}_e$ of the $^2\Sigma^+$ and $^2\Pi_{1/2}$ states are almost identical and correspond well to the theoretical predictions with a deviation of less than 5\,\%, see Table\,\ref{tab:MolePara}. The same holds for the estimated dissociation energy $\tilde{\mathcal{D}}_e$, which is  in better agreement with the values of Ref. \cite{isaev10} as therein also the low-energy part of the potentials were used to estimate the dissociation energy.

\begin{table*}[b]
	\caption{Molecular parameters of $^{226}$RaF from vibrational analysis of the electronic ground- ($X \,^2\Sigma^+$) and excited states ($A \,^2\Pi$, $B \,^2\Delta$, $C \,^2\Sigma^+$). Experimental results are compared with theoretical calculations.}
	\label{tab:MolePara} \centering
	\begin{threeparttable}
	\begin{ruledtabular}
	\begin{tabular}{lccccl}
		&$\tilde{\omega}_e$\,/$\text{cm}^{-1}$&$\tilde{T}_{e}$\,/$10^4\text{\,cm}^{-1}$&$A$\,/$10^3\text{\,cm}^{-1}$&$\tilde{\mathcal{D}}_e$\,/$10^4\text{\,cm}^{-1}$&Ref\\
			\midrule \hline
			$X \,^2\Sigma^+$& 441.8(1)&&&2.92(5)&this work\\
			&428&&&3.21&\cite{isaev13}, theo.\tnote{a}\\
			&431&&&4.26&\cite{isaev13}, theo.\tnote{b}\\
			$A \,^2\Pi_{1/2}/^2\Pi_{3/2}$& 435.5(1)/419.1(2)&1.32878(1)/1.53554(3)&2.0676(36)&2.90(3)/-&this work\\
			&432/410&1.40/1.60&2.0&3.13/-&\cite{isaev13}, theo.\tnote{a}\\	
			&428/415&1.33/1.50&1.7&&\cite{isaev13}, theo.\tnote{b}\\
			$B \,^2\Delta_{3/2}$/$^2\Delta_{5/2}$&431.9(2)/-&1.51477(2)/-&-/-&2.83(11)/-&this work\\
			&432/419&1.64/1.71&0.4&&\cite{isaev13}, theo.\tnote{a}\\	
			&431/423&1.54/1.58&0.2&&\cite{isaev13}, theo.\tnote{b}\\
			$C \,^2\Sigma$&430.9(2) &1.61806(1)&&2.78(9)&this work\\
			
			\bottomrule
		\bottomrule
	\end{tabular}%
			\begin{tablenotes}
	{
		\item [a] Fock-space coupled cluster singles and doubles (FS-CCSD), Dyall basis set and smaller active space 
		\item [b] FS-CCSD, RCC-ANO basis set and larger active space 
	}
\end{tablenotes}
\end{ruledtabular}
\end{threeparttable}
\end{table*}

%
In the case of the two low-lying $^2\Pi$ fine-structure levels, the observed origins $T_{0,0}$ agree well with the calculated values based on the RCC-ANO basis set. From the energy difference of the fine-structure components the effective spin-orbital coupling parameter $A$ is derived. For the $^2\Pi$ states the experimental value of 2068(5)\,$\text{cm}^{-1}$  is in good agreement with  the calculated value. The  band origins are in reasonable agreement with results from RCC-ANO basis set calculation,  if one attributes the $\Omega=3/2$ levels, which were
computationally found to be of mixed $\Pi_{3/2}$ and $\Delta_{3/2}$ character in this order of energies. A reverse assignment also gives better agreement with experiment.
\end{document}